# AN INVESTIGATION INTO THE EQUIVALENCY OF THREE PERFORMANCE DIMENSIONS: EVIDENCE FROM COMMERCIAL BANKS IN BANGLADESH

**Nusrat Jahan**

Adjunct Faculty, Department of Business Administration
International Islamic University Chittagong, Bangladesh
Former Assistant Professor, CIU Business School
Chittagong Independent University, Bangladesh
Email: mrs_ashfaque@yahoo.com

## ABSTRACT

This study evaluated the three dimensions of performance of commercial banks in Bangladesh by analyzing the trend of the Malmquist Productivity Index (MPI) of the Total Factor Productivity (TFP), Return on Asset (ROA) and Total Stock Return (TSR) over the period 2011 to 2015. The study developed an empirical framework with the intention to examine the equivalency of three dimensions of performance. Since, the measures of performance are different, they cannot be tested in their original form; hence, the growth rate of each category of performance measures were estimated and tested to examine the comparability among them. Evaluation of profitability revealed a decreasing trend and evaluation of stock performance suggests that investors are incurring losses on their investment over the selected period. Evaluation of productivity indicates that productivity regress was recorded initially but at the end of the studied period a modest productivity growth was recorded. Finally, this study was able to ascertain the anticipated equivalency of outcome of the three dimensions of performance.

## INTRODUCTION

Banking systems of different categories need to incorporate the ongoing changes in terms of financial deregulation, consolidation, technological advancements and financial innovation in order to remain competitive and efficient. Evaluation of different dimensions of banks' performance would suggests how competitive and efficient banks are as the industry continues to evolve. The banking sector in Bangladesh comprises of four categories of scheduled banks. These are state-owned commercial banks (SCBs), state-owned development financial institutions (DFIs), private commercial banks (PCBs) and foreign commercial banks (FCBs). Alongside the conventional interest based banking system, Bangladesh entered into an Islamic banking system in 1983. In 2015 out of the 56 commercial banks in Bangladesh, 8 operated as full-fledged Islamic banks. The increasing number of studies pertaining to the evaluation of performance of commercial banks is a result of the ongoing transformation in the financial services sector and unprecedented advancement in financial and non-financial technologies (Berger & Mester, 2003). The motivation for conducting this research stems from the fact that past literature on performance evaluation of commercial banks focused mostly on evaluating performance trends and finding the determinants of different performance measures.

However, there is no research available on the banking sector that investigated the equivalency of outcome reported by different measures of performance. This backdrop motivated the researcher to develop a research framework specifying expected equivalency of outcome for three different measures of performance. The short-term financial objective of any organization is maximization of profit over a specified period of time; and in practice, management of an organization often overlooks the long-run goal of shareholders wealth maximization and opts for profit maximization by using tools available in the industry to meet its short-term earnings target. Hence, keeping this notion in mind, profitability is considered in this study as an important dimension of performance. Theoretically, the primary financial goal of a firm is to maximize the wealth of its stockholders, which translates into maximizing the price of the firm's common stock; however, it is a long-term objective as opposed to the objective of profit maximization. Therefore, theoretical viewpoint also indicates that stock performance is an important dimension of performance. Productivity is a measure of the efficiency of production and assessing productivity is critical for any firms since all firms' are involved in converting inputs into outputs. Productivity growth is important for any firm because providing more goods and services to consumers translates into higher profits for businesses. Hence, productivity as a measure of performance presents a major interest for the regulatory authorities because an increase in the productivity level of banks contributes to the productivity level of the entire economy.

## REVIEW OF LITERATURE

The review of past literature indicates that several research exist with respect to performance evaluation of commercial banks. The following discussion details a number of studies that examined the performance of commercial banks in Bangladesh.

Sarker and Saha (2011) investigated the performance of nationalized commercial banks (NCBs), private commercial banks (PCBs), foreign commercial banks (FCBs) and state-owned commercial banks (SCBs) of Bangladesh through highlighting their profitability, branch productivity, employee productivity and overall productivity and also by using the SWOT mix during from 2000 to 2009. Rahman (2011) used the Price Earnings (PE) ratio as an indicator to measure the performance of stocks in Bangladesh. This study recommended that the stock's PE ratio should be reduced to below twenty. Uddin and Suzuki (2011) analyzed income efficiency, cost efficiency, Non-performing Loan (NPL) and Return on Asset (ROA) of 38 commercial banks in Bangladesh by applying the Data Envelopment Analysis (DEA) approach from 2001 to 2008. They found that both income efficiency and cost efficiency of all sample banks increased in 2008 compared to 2001, thus indicating improvements in bank performance during the sample period in Bangladesh. Jahan (2012) evaluated randomly selected six commercial banks in Bangladesh by using widely used indicators of banks' profitability, which are Return on Asset (ROA), Return on Equity (ROE) and Return on Deposit (ROD). This study investigated the impact of the efficiency ratio, asset utilization ratio, asset size and ROD as a determinant of banks' profitability measured by ROA. The result of regression analysis revealed that operational efficiency, asset size and ROD to be positively related and asset utilization to be negatively related to ROA, but these associations were statistically insignificant. Hoque and Rayhan (2012) applied the Data Envelopment Analysis (DEA) method to 21 commercial banks in Bangladesh and found that the scores of both input and output related technical efficiency is similar under constant returns to scale (CRS). This study revealed that the banks with higher technical efficiency possess top ranks in the banking sector. This study also suggested that continuously increasing competition in the private commercial banking sector helps to enhance the efficiency of this sector.

Haque (2013) investigated the financial performance of five private commercial banks in Bangladesh from 2006 to 2011 under four dimensions which are profitability, liquidity, credit risk and efficiency. This study found that there is no specific relationship between the generation of banks and its performance. The performance of banks is dependent more on the management's ability to formulate strategic plans and the efficient implementation of such strategies. Karim and Alam (2013) measured the financial performance of listed private sector banks in Bangladesh using three indicators. Internal-based performance measured by Return on Assets; market-based performance measured by Tobin's Q model (Price to Book ratio); and economic-based performance measured by Economic Value added. This study found that bank size, credit risk, operational efficiency and asset management have significant impact on financial performance of commercial banks in Bangladesh. Ahmed and Liza (2013) adopted Data Envelopment Analysis (DEA) method to measure efficiency of 35 commercial banks in Bangladesh from 2002 to 2011. Their study reported that most of the second and third generation banks and foreign commercial banks are highly efficient and very competitive with each other. These banks maintained not only their efficiency but also the consistency of efficiency levels from 2002 to 2011. Islam, Rahman and Hasan (2013) applied the Data Envelopment Analysis (DEA) to evaluate the efficiency of Islamic banks in Bangladesh. This study considered three inputs namely- deposits, overhead cost, total assets; and three outputs - investment and advances, Return on Investment (ROI), ROA respectively to measure efficiency. The results revealed that most of the Islamic banks were consistently efficient, both under constant returns to scale (CRS) and variable returns to scale (VRS) except Islamic Bank Bangladesh Limited (IBBL), Export Import Bank (EXIM) and Shajalal Islami Bank Limited (SIBL).

*An Investigation into the Equivalency of Three Performance Dimensions*Sufian and Kamarudin (2014) investigated efficiency and returns to scale of the banking sector in Bangladesh applying the slack-based DEA Method. They attempted to assess the level of profit efficiency of individual banks over the years 2004 to 2011. The empirical findings of their study indicated that the Bangladeshi banking sector has exhibited the highest and lowest level of profit efficiency during the year 2004 and 2011 respectively. They also found that most of the Bangladeshi banks have been experiencing economies of scale due to being at less than the optimum size, or diseconomies of scale due to being at more than the optimum size. Thus, decreasing or increasing the scale of production could result in cost savings or efficiencies. Idris (2014) investigated the technical, pure technical and scale efficiency of the Islamic banks operating in Bangladesh applying the non-parametric Data Envelopment Analysis (DEA) method. Data were analyzed in two different phases considering different input-output variables. Results of the first phase revealed that technical efficiency of all the Islamic banks were very high, amounting to an average of 98 percent, 96 percent, 98 percent and 96 percent in 2010, 2011, 2012, and 2013 respectively. Results of the second phase indicated that all the Islamic banks were technically efficient in all the periods of the study; except for 2012. Shajalal Islami Bank Limited (SIBL), Al-Arafah Islami Bank (AAIBL), and ICB Islamic banks were found to be technically inefficient.

Rahman, Hamid and Khan (2015) investigated the determinants of profitability of 25 commercial banks in Bangladesh for the period 2006 to 2013. Three different measures of profitability namely return on assets (ROA), net interest margin over total assets (NIM) and return on equity (ROE) were used in this study. The findings suggested that capital strength and loan intensity have a significant positive impact on profitability while cost efficiency and off-balance sheet activities have significant negative impact on banks profitability. Non-interest income, credit risk and growth of GDP were found to be important determinants of NIM. Rahman (2016) provided a comprehensive financial performance evaluation of scheduled commercial banks in Bangladesh in his study. This study revealed that most of the banks except a few have shown poor economic performance, negative economic value added (EVA) and undervalued market price per share. The research findings have also revealed that there is significant correlation between EVA and Profitability Ratio and EVA and Market Ratio. Banna, Ahmad and Koh (2017) examined the effect of the global financial crisis

73

and other factors on the efficiency of commercial banks in Bangladesh by employing the Data Envelopment Analysis (DEA) method. Their findings indicated that the banking sector has exhibited the highest efficiency level during 2001, while efficiency seemed to be at the lowest level during 2010. This study found that crisis along with bank size, capital adequacy ratio, return on average equity and real interest rate has a significant effect on bank efficiency. Islam, Sarker, Rahman, Sultana and Prodhan (2017) examined the profitability determinants of private commercial banks in Bangladesh for the years 2014 and 2015. Their study reported that asset size and the Net Interest Margin ratio had no significant effect on profitability but the effect of non-performing loans to total loans (NPL) on profitability was observed as the most significant among various variables. This study indicated that investment activities, mainly in shares and debentures of private sectors have a positive impact on return on equity (ROE). Robin, Salim and Bloch (2018) examined the cost efficiency of the commercial banks in Bangladesh in the context of financial reform by employing the Single-stage Stochastic Frontier Analysis (SFA) model. Their research reported that banks costs have fallen due to financial deregulation and also suggest that the presence of politically linked directors in the bank board has an adverse effect on efficiency.

This paper aims to expand the established literature on bank performance by examining the profitability, productivity and stock performance of commercial banks in Bangladesh altogether. A majority of the studies on bank performance in Bangladesh have concentrated on determinants of profitability but studies comparing profitability, productivity and stock performance of both Islamic banks and conventional banks are scarce. Only a limited number of studies have attempted to investigate the notion of shareholders wealth maximization as a criterion for evaluating performance of commercial banks which is surprising, given that creating value for shareholders has been the main strategic objective of quoted banks over the last decades. Studies focusing on shareholders' value creation by using indicators such as Tobin's Q, market value added, market value to book value, price to earnings (PE) ratio, market adjusted returns (MAR), economic value added (EVA) to invested capital and stock return, are characterized by their forward-looking aspect and reflection of shareholders' expectation concerning the firm's future performance, which has its basis on previous or current performance (Wahla, Shah & Hussain, 2012; Shan

& McIver 2011; Ganguli & Agrawal, 2009). In this study, total stock return (TSR) was used as a metric to represent the gains in shareholders' wealth which is the best measure of value creation according to Brealey and Myers (1991). There are substantial studies performed with regard to the efficiency and productivity of financial institutions in developed countries, however, empirical evidences on the banking sector of Bangladesh are limited. Review of past literature reveals that empirical evidence on the application of Malmquist Productivity Index (MPI) for productivity analysis of both conventional and Islamic banks of Bangladesh were scarce. Finally, no research in the context of the banking sector in Bangladesh or otherwise were found to develop a framework and validate the equivalency of outcome generated by measures of productivity, profitability and stock performance.

## OBJECTIVES OF THE STUDY

The main objectives of this study are as follows:

1. First, to evaluate the trend of commercial banks' performance by applying three different measures which are productivity, profitability and stock performance.

2. Second, to identify which banking system is boasting superiority in terms of productivity, profitability and stock performance.

3. Third, to examine the equivalency of outcome generated by the three dimensions of performance.

## RESEARCH FRAMEWORK

Evaluation of firm's performance can offer significant invaluable information in assisting management to monitor performance, report progress, improve motivation and communication, and pinpoint problems (Waggoner, Neely & Kennerley, 1999). Although there are a wide variety of evaluation criterion brought forward by past research to assess the financial performance of a firm, in this study, evaluation criterions are categorized into productivity performance, profitability performance and stock performance. The researcher developed the following framework depicted in Figure 1 to support the classification and used it as a basis for evaluating the performance of commercial banks in terms of these three dimensions.

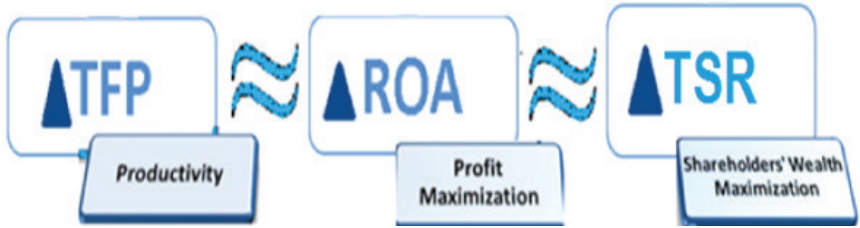

**Figure 1: Research Framework**

The researcher states that the main goal of any firm is to convert inputs into outputs. Hence, the natural measure of performance for any firm is evaluation of its productivity which is generally a ratio of outputs to inputs, where larger value of this ratio is associated with better performance. Productivity growth of a firm stems from its managerial efficiency, scale efficiency and adaption to changing technology, all of which are encompassed and reflected through the measure of the Total Factor Productivity (TFP). The researcher recognizes that profit maximization is considered as the major financial goal of any firm in the short-run. Firm's stakeholders are more concerned about the profitability of a firm while making any fiscal decision regarding that firm. The current financial literature suggests that the main financial goal of a firm is maximization of its shareholders' wealth and it is generally considered as a long-term financial goal. Shareholders' wealth creation is possible through stock price maximization and increased profitability is a prerequisite for stock price maximization along with other factors in place. This study expects these three dimensions of performance to generate comparable result indicating equivalent percentage of progress and regress in all three dimensions. Hence, change in the Total Factor Productivity (TFP) is expected to be equivalent to change in Return on Asset (ROA) and change in Return on Asset (ROA) is expected to be equivalent to change in Total Stock Return (TSR). On the basis of these arguments, the researcher conducted this study to make this framework functional by examining the association anticipated in the model.

## RESEARCH METHODOLOGY

### Sample, Sources and Type of Data

There are 56 commercial banks in Bangladesh of which 48 are conventional and 8 are Islamic banks. The sample comprises of these 29 listed commercial banks, of which 23 are conventional and 6 are Islamic banks and the data collected covers the time period between 2011 and 2015. The listed commercial banks are selected because only these banks can supply data on stock prices required for stock performance. For calculation of the Malmquist Productivity Index (MPI) of the Total Factor Productivity (TFP), the values of input and output variables were obtained from the year 2010 to 2016 to constitute a balanced panel database. The data was obtained from audited annual reports, annual reports of Bangladesh Banks and also the financial stability report published by the Bangladesh Bank. The nature of this dataset is pooled time-series cross-sectional (TSCS) data where pooled arrays of data were combined in cross-sectional data on 'n' spatial units (i.e. 29 banks) and 't' time periods (i.e. 5 year time periods) to produce a data set of 'n × t' (i.e. 29 X 5 = 145) observations for this study. This study employed a balanced panel database of input and output variables for estimation of the Malmquist Productivity Index (MPI) of the Total Factor Productivity (TFP).

### Measure of Profitability, Productivity and Stock Performance

In line with past studies, Return on Asset (ROA) is used as a measure of profitability which indicates the profit earned per dollar of assets and reflects how well bank management uses the bank's real investments resources to generate profits (Naceur, 2003; Alkassim, 2005). Productivity of a firm is a ratio of output to inputs and when the production process involves more than one input and one output which is often the case, then a method for aggregating these inputs into a single index of inputs and a single index of outputs must be used to obtain a ratio measure of productivity. Coelli et al. (1998) termed Total Factor Productivity (TFP) as an overall productivity measure that encompasses the productivity of all production factors and outputs. In conformity with past studies, this study utilized the index number approaches to estimate productivity change and employed the non-parametric Malmquist (1953) Productivity Index (MPI) of Caves,

Christensen and Diewer (1982) to estimate the Total Factor Productivity Change Index (TFP) of commercial banks. To derive the MPI through an output orientated Data Envelopment Analysis (DEA) model, data on inputs and outputs were fed into the intermediation approach of modelling bank behaviour. This study employed input and output variables of Batchelor (2005) for estimation of the Malmquist Productivity Index of the Total Factor Productivity. Chu and Lim (1998) used end of the year stock prices whereas Becalli, Casu and Girardone (2006) calculated the annual returns by adding daily returns. However, this study estimated annual total return on stock by taking account of both capital gain and cash dividend earned by the shareholders. The Total Stock Return (TSR), in its simplest form, is calculated as the increase in market capitalization for a period plus the total dividend paid in the period, expressed as a percentage of the market capitalization at the start of the period. According to Brigham and Ehrahdat (2009):

**Annual Total Stock Return** = Capital Gain Yield + Dividend Yield

$$\textbf{Annual Total Stock Return} = \frac{P_1 - P_0}{P_0} + \frac{D_1}{P_0}$$

## Statistical Tools and Programs Used for Evaluating Performance

The data collected for performance evaluation of commercial banks were tabulated and descriptive measures such as trend analysis, average, cumulative average, maximum, minimum, standard deviation, coefficient of variation and yearly growth rate were computed using Excel. The measures of inferential statistics which are one way ANOVA and paired sample 't' test were computed using the SPSS. Malmquist Productivity (MPI) of Total Factor Productivity (TFP) change index was estimated by employing a balanced panel database of input and output variables in the DEAP version 2.1 developed by Coelli (1996).

## Hypothesis Formulation for Examining the Equivalency of Outcome

### *Hypothesis I*

The study formulated the following null hypothesis to evaluate the equivalency of the Mean Growth Rate of Profitability, Stock return and Productivity.

**H1o:** There is no difference among mean growth rate of profitability, stock return and productivity.

The three measures of performance dimension used in this study generated outcome which are non-comparable. Since the measures of profitability (ROA) and stock return (TSR) are financial ratios whereas the productivity measured by the Total Factor Productivity (TFP) index generates a score. Hence, to make all three dimensions comparable the yearly growth rate of these three measures of performance was calculated for all listed commercial banks. To test the first null hypothesis, one-way ANOVA is carried out to check if any significant differences exists among the mean growth rate of these three measures of performance. If the reported value of 'P' for "F" test is less than the 0.05 level of significance the first null hypothesis (H1o) will be rejected at the 5% level of significance.

### *Hypothesis II, III and IV*

The study formulated the following null hypotheses to evaluate the differences that arise between mean growth rates of two performance measures.

**H2o:** There is no difference between the mean growth rate of the ROA and the mean growth rate of the TSR.
**H3o:** There is no difference between the mean growth rate of the TSR and the mean growth rate of the TFP.
**H4o:** There is no difference between the mean growth rate of the ROA and the mean growth rate of the TFP.

One way ANOVA only indicates if the mean difference between two or more groups is statistically significant or not. To identify the differences that arise between the means of two groups, the three dimensions of performance are paired into groups of two to identify the numerical mean difference between them and also to test if the difference is statistically significant or not. Hence, the paired-sample 't' test was conducted to test the second, third and fourth null hypothesis. If the reported value of 'P' is less than the 0.05 level of significance for each of these three hypotheses separately, then the null hypotheses (H2o, H3o, and H4o) will be rejected at the 5% level of significance.

# FINDINGS AND ANALYSIS

## Descriptive Statistics for Three Dimensions of Performance

The following Table 1 reports the overall mean, standard deviation, maximum and minimum values of the Total Factor Productivity (TFP), ROA and the Total Stock Return (TSR). The table presents the breakdown of variables into a between and within variations. Table 1 reports that total number of observation N=145, number of banks n=29 and time period t=5 (i.e. 2011-2015). The overall was calculated for 145 observations, between values were calculated across 29 banks and within values were calculated across the selected time period for each bank. In this table 'bn' indicates bank's name coded with numerical values 1 to 29 and 'bc' indicates category of banks, also coded with numerical values; 1 for conventional and 2 for Islamic banks and the year indicates the selected period of study.

**Table 1: Descriptive Statistics for Three Dimensions of Performance**

| Variable | | Mean | Std.Dev. | Min. | Max | Observations |
|---|---|---|---|---|---|---|
| ROA | Overall | 0.122132 | 0.0074489 | 0.0008 | 0.07 | N =145 |
|  | Between |  | 0.0045476 | 0.0038 | 0.02444 | n = 29 |
|  | within |  | 0.0059481 | -0.0026268 | 0.0577732 | T = 5 |
| TSR | Overall | -0.26176 | 0.3337158 | -0.96 | 0.44 | N =145 |
|  | Between |  | 0.0859483 | -0.408 | -0.054 | n = 29 |
|  | within |  | 0.32227759 | -1.088359 | 0.4202414 | T = 5 |
| TFP | Overall | 0.980689 | 0.1953504 | 0.323 | 2.993 | N =145 |
| | Between | | 0.0541764 | 0.9044 | 1.2282 | n = 29 |
| | within | | 0.1879048 | 0.0754896 | 2.74549 | T = 5 |
| bn | Overall | 15 | 8.395601 | 1 | 29 | N =145 |
| | Between | | 8.514693 | 1 | 29 | n = 29 |
| | within | | 0 | 15 | 15 | T = 5 |
| bc | Overall | 1.206897 | 0.4064848 | 1 | 2 | N =145 |
| | Between | | 0.4122508 | 1 | 2 | n = 29 |
| | within | | 0 | 1.206897 | 1.206897 | T = 5 |
| year | Overall | 2013 | 1.419119 | 2011 | 2015 | N =145 |
| | Between | | 0 | 2013 | 2013 | n = 29 |
| | within | | 1.419119 | 2011 | 2015 | T = 5 |

## Trend Analysis of Profitability of Commercial Banks

The evaluation of bank's profitability is done by trend analysis of the ROA. Table 2 and Figure 2 present the trend in yearly ROA for a five year period. Table 2 also reports the five year cumulative average and variation in yearly ROA of conventional banks and Islamic banks by standard deviation and coefficient of variation. The maximum and minimum values of yearly ROA and the standard deviation and coefficient of variation of yearly average ROA of all listed commercial banks are also reported in Table 2.

Table 2: Trend Analysis of Profitability

| Banks | 2011 | 2012 | 2013 | 2014 | 2015 | Five Year's Average | SD | CV |
|---|---|---|---|---|---|---|---|---|
| Conv. | 0.017 | 0.0117 | 0.0099 | 0.0105 | 0.0099 | 0.012 | 0.003 | 0.2541 |
| Islamic | 0.0179 | 0.0148 | 0.0109 | 0.0104 | 0.0097 | 0.013 | 0.004 | 0.2747 |
| **All Banks** | | | | | | | | |
| **Max** | 0.0401 | 0.07 | 0.0168 | 0.0236 | 0.0208 | | | |
| **Min** | 0.0061 | 0.001 | 0.0019 | 0.001 | 0.0008 | | | |
| **Yearly Average** | 0.017 | 0.012 | 0.0101 | 0.0105 | 0.0099 | 0.012 | 0.003 | 0.2548 |
| **SD** | 0.007 | 0.012 | 0.0038 | 0.0048 | 0.0045 | | | |
| **CV** | 0.408 | 0.996 | 0.379 | 0.4533 | 0.4537 | | | |

The trend in ROA indicates that average ROA of both conventional and Islamic banks remained higher than average ROA of all listed commercial banks initially; however, profitability of both types of banks reported a declining trend over the studied period. The result shows that Islamic banks five years average ROA, which is 1.3%, is higher than the total listed banks' five years average ROA of 1.2% However, the Conventional banks and all listed banks' five years average ROA is identical, that is 1.2%.

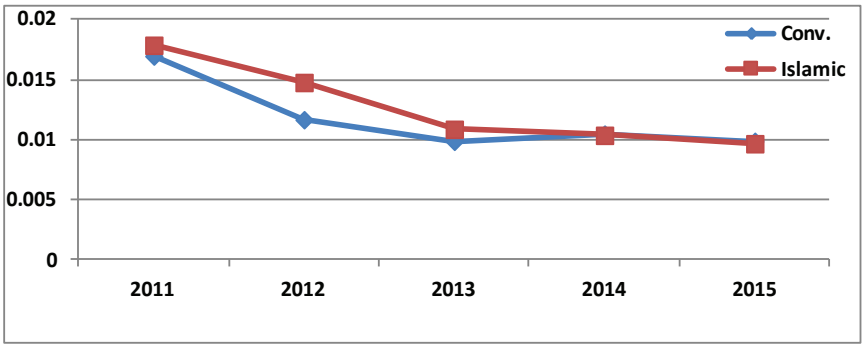

**Figure 2: Trend Analysis of Profitability**

Furthermore, in Table 2, the variation reported by standard deviation of 0.3% and coefficient of variation of 25.4% are identical for both conventional banks and all listed commercial banks. Though, Islamic banks five years average ROA is higher than conventional banks and all listed commercial banks but the variation from five years average ROA as reported by standard deviation of the ROA is higher for Islamic banks compared to conventional banks and all listed commercial banks. In addition, the coefficient of variation of 27.5% indicates that year to year variation of the ROA of Islamic banks is also high compared to conventional banks and all listed banks. Therefore, profitability of Islamic banks is more volatile compared to conventional banks. Overall, the trend analysis of the ROA and comparison of the ROA among banking groups and all listed commercial banks shows that both banking groups reported to have above average profitability over the studied period.

## Trend Analysis of Stock Performance of Commercial Banks

This study evaluated the stock performance of conventional and Islamic banks by the annual total stock return (TSR) which is the sum of capital gain yield and dividend yield. The trend in yearly TSR for a five year period is presented in both Table 3 and Figure 3. In Table 3, five years cumulative average and variation in yearly TSR of Conventional banks and Islamic banks by standard deviation and coefficient of variation are also reported. Table 3 also presents the maximum and minimum values of yearly TSR and the standard deviation and coefficient of variation of yearly average TSR of all listed commercial banks in Bangladesh.

**Table 3: Trend Analysis of Stock Performance**

| Banks | 2011 | 2012 | 2013 | 2014 | 2015 | Five Year's Average | SD | CV |
|---|---|---|---|---|---|---|---|---|
| Conv. | -0.7987 | -0.4191 | -0.1541 | -0.0252 | -0.0087 | -0.2812 | 0.3327 | 1.1833 |
| Islamic | -0.5483 | -0.2567 | -0.1333 | -0.1367 | 0.1283 | -0.1893 | 0.2451 | 1.2942 |
| All Banks | | | | | | | | |
| Max | -0.34 | -0.13 | -0.41 | 0.23 | 0.44 | | | |
| **Min** | -0.96 | -0.68 | -0.47 | -0.28 | -0.32 | | | |
| Yearly Average | -0.7469 | -0.3855 | -0.1498 | -0.0482 | 0.0196 | -0.2622 | 0.3114 | -1.1879 |
| SD | 0.2389 | 0.1303 | 0.1807 | 0.1539 | 0.199 | | | |
| **CV** | -0.3199 | -0.3379 | 0.0016 | -3.191 | 10.18 | | | |

The results presented in Table 3 and Figure 3 reports that yearly average TSR across the five year period are mostly negative due to the stock market crisis of 2009, thus indicating annual average losses for both the banking group and all the listed commercial banks. The result revealed that losses incurred by investors of Islamic banks stocks remain lower than the losses borne by investors of conventional banks till 2013. However, the losses incurred by investors were higher on the stock of Islamic banks compared to conventional banks and all listed commercial banks during the year 2014. Though, the losses sustained by investors of Islamic banks showed a declining trend with a reported increase in losses in 2014 the return on stock of Islamic banks gained a momentum during 2015, boasting a positive TSR of 12.83%.

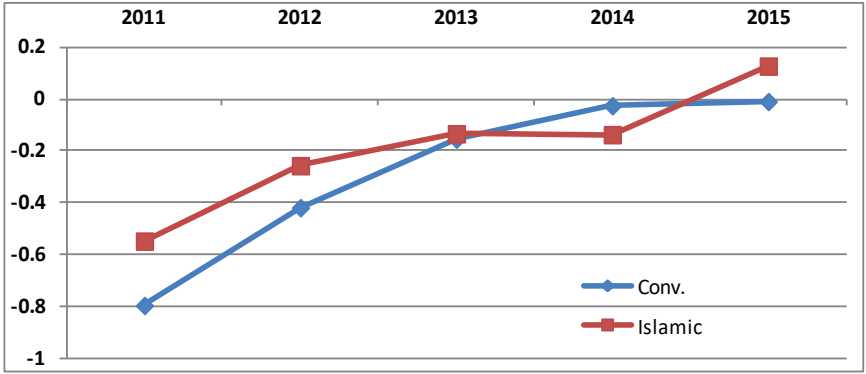

**Figure 3: Trend Analysis of Stock Performance**

From Table 3, it is visible that the losses on the stock of conventional banks and also the TSR of all listed commercial banks began reducing from the year 2011 with a greater reduction in losses during 2014. In 2015, the TSR of all commercial banks became positive as the stock market started to show signs of recovery due to the different corrective actions taken by the authorities after the stock market crash of 2010. Table 3 also reports that Islamic banks' five years cumulative average TSR, which is -18.93%, is lower than the total listed banks' and conventional banks cumulative TSR of -26.22% and -28.12% respectively. However, the variation recorded in the year to year TSR by coefficient of variation is higher for Islamic banks i.e. 129.42% compared to conventional and all listed commercial banks. Though, the conventional banks' five years cumulative average TSR remained higher than all listed commercial banks but the variation of yearly average TSR of both conventional banks and all listed banks remained identical which is 118%. Therefore, though the losses incurred from investment in the stock of Islamic banks remained low over the study period but the volatility of stock performance of Islamic banks was more compared to all listed commercial banks.

## Trend Analysis of Productivity of Commercial Banks

The trend analysis of the MPI of Total Factor Productivity (TFP) of conventional and Islamic banks over the period 2011 to 2015 is presented in Table 4 and Figure 4. Table 4 also presents the five year cumulative average and variation in the yearly TFP change index of Conventional banks and Islamic banks by standard deviation and coefficient of variation. The maximum and minimum values of the yearly TFP change index and the standard deviation and coefficient of variation of yearly average TFP change index of all listed commercial banks are also reported in Table 4.

**Table 4: Trend Analysis of Total Factor Productivity Change Index**

| TFP Change | 2010-2011 | 2011-2012 | 2012-2013 | 2013-2014 | 2014-2015 | Five Year's Average | SD | CV |
|---|---|---|---|---|---|---|---|---|
| Conv. | 0.90020 | 0.93248 | 0.96539 | 0.99670 | 1.09830 | 0.97861 | 0.07600 | 0.07766 |
| Islamic | 0.99420 | 1.02467 | 0.92833 | 1.00667 | 0.98950 | 0.98867 | 0.03637 | 0.03679 |
| **All Banks** | | | | | | | | |
| **Max** | 1.15300 | 1.11100 | 1.049 | 1.11700 | 2.99300 | | | |
| Min | 0.61900 | 0.67100 | 0.825 | 0.32300 | 0.89700 | | | |
| Yearly Average | 0.91962 | 0.95155 | 0.9577 | 0.99876 | 1.07579 | 0.98069 | 0.06016 | 0.06134 |
| SD | 0.11476 | 0.07963 | 0.05218 | 0.13979 | 0.37349 | | | |
| CV | 0.12479 | 0.08368 | 0.00046 | 0.13997 | 0.34717 | | | |

The trend of the Total Factor Productivity change index of Islamic banks indicated that it is failing to maintain the progress made in some years and thus reporting a score of less than 1 during 2011, 2013 and 2015. But the trend of the Total Factor Productivity change index of the conventional banks though remained in regress during the year 2011 to 2014 but the trend was consistent and rising; and finally reported a progress during 2015.

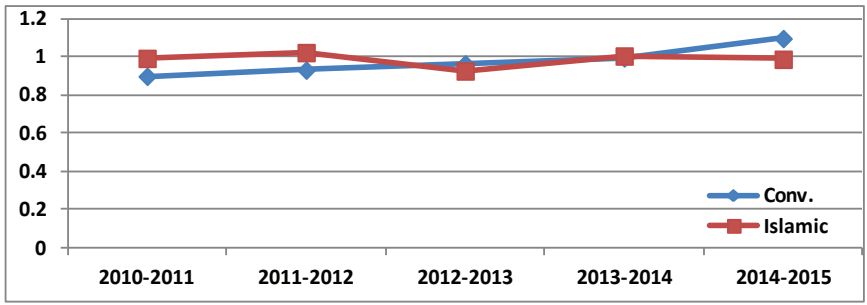

**Figure 4: Trend Analysis of Total Factor Productivity Change Index**

On an average, productivity regress was recorded initially for all listed commercial banks and the conventional banking group but at the end of the studied period all listed banks recovered and recorded a modest productivity growth of 1.07579 during 2015. Total Factor Productivity change Index of Islamic banking operations' reports that productivity growth was more dominant relative to conventional banks during 2012 and 2014. Though, the TFP change index was found to be higher for Islamic banks but conventional banks' reported a steady increment and reported a productivity progress at the end of the studied period. Evaluation of productivity performance of commercial banks in Bangladesh revealed that Islamic banks, on an average, had a relatively higher five year cumulative average Total Factor Productivity change index (0.988) compared to that of conventional banks (0.9786) over the studied period. However, for both the banking groups the reported five year cumulative mean indicated an overall productivity regress. In addition, Islamic banks reported less variability in its TFP change index with lower CV (0.03679) compared to CV (0.07766) of conventional banks.

## Equivalency of Mean Growth Rate of Profitability, Stock return and Productivity

The research framework developed in this study anticipated that the outcome generated by three dimensions of performance may perhaps be equivalent. The researcher anticipated that the rate of progress or regress in one dimension could be equivalent to the rate of progress or regress of other dimensions of performance. That means, if one dimension indicates a positive growth the other two dimensions may also indicate the same. However, the measurement criterions of these three measures of performance were not same. The measures of profitability (ROA) and stock return (TSR) are financial ratios whereas the productivity measured by the Total Factor Productivity (TFP) generates a score. Hence, to make all three dimensions comparable the yearly growth rate of these three measures of performance was calculated for all listed commercial banks. Subsequently, to examine the equivalency of these three measures, one-way ANOVA is carried out to check if there is any significant difference among the mean growth rate of these three measures of performance.

**Table 5: One Was ANOVA Test for Equivalency of Mean Growth Rate**

|  | Sum of Squares | df | Mean Square | F | Sig |
|---|---|---|---|---|---|
| **Between Groups** | 27.852 | 2 | 13.791 | 0.503 | 0.605 |
| **Within Groups** | 11681.831 | 426 | 27.422 |  |  |
| Total | 11709.413 | 428 |  |  |  |

Table 5 presents the output of the ANOVA analysis. This table reports that the result of the "F" test is found to be statistically insignificant since the reported value of 'P' is high above the 0.05 level of significance. Hence, this study failed to reject the first null hypothesis (H1o) that states there is no statistically significant difference among the mean growth rate of the three measures of performance. Therefore, this finding suggests that these three dimensions of performance - profitability, stock return and productivity are expected to generate equivalent outcome for commercial banks in Bangladesh. Thus, if one dimension reports improvement in performance of commercial banks, the other two dimensions also evidently reported progress in performance and vice-versa. One way ANOVA only indicates if the mean difference between two or more groups is statistically significant or not. However, by conducting the paired-sample 't' test, it is possible to identify the differences that arise between the means of two groups and the result of which is reported in Table 6.

**Table 6: Paired-Sample 't' Test for Equivalency of Mean Growth Rate**

|  | Mean | Std. Deviation | Std. Error Mean | 95% Confidence Interval of the Difference | | t | df | Sig. (2-tailed) |
|---|---|---|---|---|---|---|---|---|
|  |  |  |  | Lower | Upper |  |  |  |
| Pair 1 ChROA-ChTSR | -0.32251 | 7.6140805 | 0.7069496 | -1.75258 | 1.048080 | -0.498 | 115 | 0.619 |
| Pair 2 ChTSR-ChTFP | 0.235619 | 7.6804473 | 0.7131116 | -1.17692 | 1.648156 | 0.330 | 115 | 0.742 |
| Pair 3 ChROA-ChTFP | -0.11663 | 0.9068722 | 0.0842009 | -0.28341 | 0.050154 | -1.385 | 115 | 0.169 |

Hence, the three dimensions of performance were paired into groups of two to identify the numerical mean difference between them and also to test if the difference is statistically significant or not. On an average, percentage change in the ROA were -0.352251 unit lower than the percentage change in the TSR but the difference between the percentage change in the ROA and the percentage change in the TSR is not statistically significant at the 5% level of significance since the reported 'P' value is high above 0.05. Therefore, this study accepts the second null hypothesis (H2o) that states there is no significant difference between the mean growth rate of the ROA and the mean growth rate of the TSR. Furthermore, percentage change in the TSR were on average 0.23561902 unit higher than the percentage change in the TFP but the difference between the percentage change in the TSR and percentage change in the TFP is not statistically significant at the 5% level of significance since the reported 'P' value is high above 0.05. Hence, this study also fails to reject the third null hypothesis (H3o) and concludes that there is no significant difference between the mean growth rate of the TSR and the mean growth rate of the TFP. Percentage change in the ROA were -0.116632 unit lower than the percentage change in the TFP but the difference between the percentage change in the ROA and the percentage change in the TFP is not statistically significant at the 5% level of significance since the reported 'P' value is high above 0.05. Finally, this study also failed to reject the fourth null hypothesis (H4o) and concludes that there is no significant difference between the mean growth rate of the ROA and mean growth rate of the TFP. Thus, the mean growth rate of the ROA and the TSR, the mean growth rate of the TSR and the TFP; and the mean growth rate of the ROA and the TFP evidently regress and progress almost in a parallel manner for all listed commercial banks in Bangladesh.

## CONCLUSION

In this paper, commercial banks in Bangladesh were evaluated in terms three dimensions of performance which are profitability, stock performance and productivity performance. This study brings insights into the banking literature in Bangladesh by evaluating stock performance with total stock return and productivity performance with the Malmquist Productivity Index (MPI) of the Total Factor Productivity. The novel feature of this study is the development of an empirical framework that examines the equivalency of three dimensions of performance to determine whether these performance indicators generate comparable results. Since, the measures of performance are different, they cannot be tested in their original form; hence, growth rate for each category of performance measures were estimated and tested to examine the comparability among them. Evaluation of profitability of listed

*An Investigation into the Equivalency of Three Performance Dimensions*

commercial banks in Bangladesh revealed that on an average, profitability is exhibiting a decreasing trend over the selected period. The profitability performance of Islamic banks was rather high but volatile compared to conventional banks. Islamic banks operate with smaller network of branches especially in urban areas in Bangladesh, thus allowing them to control and coordinate operations more effectively.

This study employed the time frame where the stock return of all listed companies was largely affected by the disastrous stock market collapse of 2010. Therefore, in this study, analysis of stock performance of listed commercial banks in Bangladesh indicated that investors are incurring losses on their investment on stocks of both types of banks, mostly due to lack of investors' confidence in the stock market thus affecting both demand and the price of stock. However, the trend analysis reported that the losses in terms of total stock return are declining and report a positive return at the end of the selected period. The trend in return of all listed commercial banks and conventional banks are somewhat similar.

However, the relative losses from investment in stock are lower in Islamic banks compared to conventional banks. In addition, conventional banks failed to report any positive return at the end of the selected period unlike Islamic banks. On the whole, it is evident that investment in stock of commercial banks in Bangladesh is becoming gainful at the end of the studied period due to the undertaking of different systematic measures like demutualization of stock exchanges, implementation of merchant bankers and portfolio manager regulation, holding of 30% and 2% of shares of the listed company by sponsors and directors respectively. These initiatives are reinstating public faith in the stock market, thereby stabilizing the capital market in the country. Evaluation of productivity performance of commercial banks clearly indicates that the TFP change index has exhibited a marginal progress over the selected years. This finding suggests that commercial banks in Bangladesh are presumably falling behind in productivity due to their delayed compliance to technological developments whilst financial innovation assisted by technological advancements are globally transforming the role of financial intermediaries and also intensifying competition. Hence, to remain competitive commercial banks in Bangladesh must respond to technological changes at the same pace they are adapted globally by engaging in financial innovation assisted by technology and also offering new financial solutions to customers for better investment opportunities, and returns and risk management.

This research has provided evidence that the three dimensions of performance generates comparable performance results for commercial banks in Bangladesh. Hence, it is expected that the policies and directives of the Bangladesh Bank would be directed towards enhancing the efficiency and productivity of banks with the aim of achieving major financial goals of profit maximization and shareholders wealth maximization in order to build resilience and retain stability in the country's financial system.